\documentclass[aps,showpacs,amssymb,12pt,prb]{revtex4}
\usepackage{graphicx,graphics}
\begin{document}
\title{Drag of ballistic electrons by an ion beam
}
\author{V. L. Gurevich and M. I. Muradov}
\affiliation{Ioffe Institute, Russian Academy of Sciences,
194021 Saint Petersburg, Russia}
\vspace{0.7truecm}
\begin{abstract}
\baselineskip=2.5ex \indent Drag of electrons of 1D ballistic
nanowire by a nearby 1D beam of ions is considered. We
assume that the ion beam is represented by an ensemble of heavy
ions of the same velocity $\bf V$. The ratio of the drag current to
primary current carried by the ion beam is calculated.

The drag current appears to be a nonmonotonic function of velocity
$V$, it has maxima for $V$ near $v_{nF}/2$ where $n$ is the number
of electron miniband (channel) and $v_{nF}$ is the corresponding
Fermi velocity. This means that the ion beam drag can be applied for
ballistic nanostructure spectroscopy.
\end{abstract}

\maketitle \baselineskip=3.5ex

\section{Formulation of the problem}
Drag as a physical phenomenon in solids can be described as follows.
Consider a solid with two types of quasiparticles (type 1 and type 2). One creates
a flux of the quasiparticles of type 2, the so-called {\em driving current}.
As a result of the interaction between particles a current of quasiparticles of type 1, the
so-called {\em drag current} is excited. An example of such phenomenon is the Coulomb drag
where a current in conductor creates a current in adjacent conductor --- see papers
by Pogrebinski~\cite{Pog} and Price~\cite{Price}.

The purpose of the present paper is to consider a somewhat
different situation where the driving current is created by real
particles outside the conductor (rather than by quasiparticles
within it). This would provide a {\em contactless method} to
generate a drag current (or voltage) in a nanostructure.

Two formulations of the problem are feasible.
\begin{enumerate}
\item The dragging flux consists of heavy ions of almost the
same velocity $\bf V$.

\item One considers a flux of weakly ionized gas that is in thermal
equilibrium having some temperature $T$ and hydrodynamical
velocity $\bf V$.
\end{enumerate}

In the present paper we treat the first possibility. In other words,
we consider an ion beam, i. e. a flux of ions having the same velocity
$\bf V$. For the simplest situation the value of velocity $\bf V$ is
determined by the accelerating voltage $\mathfrak{V}$ and the
ion mass $M$ as
\begin{equation}
\frac{MV^2}2=e_I\mathfrak{V}.
\label{1}
\end{equation}
Here $e_I$ is the charge of an ion.

For the drag system we will treat a ballistic (collisionless) electron transport in a
quantum wire. Such nanoscale systems may have rather low electron densities that can be
varied by means of the gate voltage. The collisionless quantum wires act as waveguides
for the electron de Broglie waves. For a strong Fermi degeneracy
\begin{equation}
T\ll\mu
\label{2a}
\end{equation}
where $T$ is the temperature (we will use the energy units for it) while $\mu$ is
the Fermi energy, each miniband of transverse quantisation (channel) gives the following
contribution to the conductance
\begin{equation}
G_0=\frac{e^2}{\pi\hbar},
\label{2}
\end{equation}
($e$ being the electron charge) so that the total conductance is
$$
G={\cal N}G_0.
$$
Here ${\cal N}$ is the number of such active channels,
i. e. the minibands with bottoms $\epsilon_n(0)$ below the Fermi level $\mu$.

Our purpose is to investigate the main features of this drag
phenomenon. We assume that the distance $d$ between the ion beam
and the wire is much larger than the width of the wire, so that on
the scale of this width the Coulomb interaction of ions and
electrons is a smooth function. Then the selection rules for
the corresponding matrix elements require that electrons involved
in the transitions change their quasimomenta but remain within
the initial transverse quantized channel
$n$. One can vary the velocity $\bf V$ of the ions with the
accelerating voltage $\mathfrak{V}$ and measure the resulting
variation of the drag current (or drag voltage). We will denote by
${\cal V}_{\bf r}$ the volume occupied by the nanowire while
${\cal V}_{\bf R}$ will be the volume where the flux of ions
propagates and interacts with the electrons of the nanowire. We
assume both ${\cal V}_{\bf r}$ and ${\cal V}_{\bf R}$ to have a 1D
shape of length $L$ parallel to $z$-axis.

\begin{figure}[htb]\label{perechod}
\begin{center}
\includegraphics[width=5 cm]{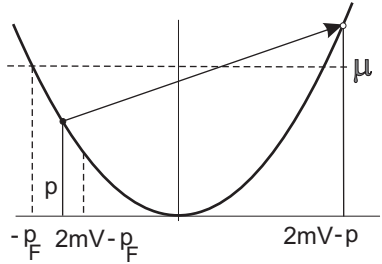}\\
\end{center}
\caption{Momenta from $-p_F$ to $2mV-p_F$ are involved in
transitions. If $V>v_F/2$ all negative momenta $p$ contribute to
the drag current.}
\end{figure}

One can give the following qualitative considerations concerning the drag by
an ion beam. Due to the conservation of such quantities as the energy, the transverse quantized
channel number $n$ and the (quasi)momentum in the electron-ion collisions one has to consider
in the Born approximation the transition of electron from $|n,p\rangle$
to $|n,p+q_z\rangle$ state (where $p$ is the $z$-component of electron quasimomentum)
and that of the ion from $|{\bf P}\rangle$ to
$|{\bf P}-{\bf q}\rangle$ state according to relation
\begin{equation}\label{qc01}
\frac{p^2}{2m}+\frac{{\bf
P}^2}{2M}=\frac{(p+q_z)^2}{2m}+\frac{({\bf P}-{\bf q})^2}{2M}.
\end{equation}
The $\delta$-function describing the energy conservation can be therefore
written as
\begin{equation}\label{qc02}
\delta\left[\frac{q_z^2}{2m}(1+m/M)+\frac{q_z}{m}(p-mV)+\frac{q_{\perp}^2}{2M}\right]\simeq\,
\frac{2m}{|q_z|}\delta\left[q_z-2(mV-p)\right],
\end{equation}
where $P_z\equiv P=MV$. Further on we will take into account that $m/M\,\ll\,1$ and
neglect $m/M$ as compared to 1 and $(m/M)q_{\perp}^2$ as compared to $q_z^2$.
Therefore the transferred (quasi)momentum is $q_z=2(mV-p)$ and the
probability of such a transition includes the factor
\begin{equation}\label{qc03}
f_{np}(1-f_{n,p+q_z})-f_{n,p+q_z}(1-f_{np})=f_{np}-f_{n,2mV-p}
\end{equation}
as well as the electron-ion Coulomb interaction matrix element
squared. For the 1D situation under consideration it
has a factor proportional to
\begin{equation}\label{qc04}
\left.K_0^2(|q_z|d/\hbar)\right|_{q_z=2(mV-p)}
\end{equation}
where $d$ is the distance between the ion beam and the wire and
$K_0$ is the McDonald function [see below Eq.~(\ref{8})]. One can use for it
the following approximate equations
\begin{eqnarray}
K_0(s)\approx\ln\frac{2}{\gamma s}\quad\mbox{for}\quad s\ll1,\label{11e}\\
K_0(s)\approx \sqrt{\frac{\pi}{2s}}e^{-s}\quad\mbox{for}\quad
s\gg1 \label{11f}
\end{eqnarray}
where $\ln{\gamma}=0.5772$. The
drag current is proportional to the sum over electron momenta $p$
of the products in Eqs.~(\ref{qc02}) --- (\ref{qc04}). We consider $p<0$
and require the state $p$ to be occupied, this condition
leads to $-p_F<p<0$. The requirement that the final state with momentum
$2mV-p$ should be empty gives $2mV-p>p_F$ provided $V<v_F/2$ (see
Fig.~\ref{perechod}). If $V>v_F/2$ there is no additional
restriction except $-p_F<p<0$, i.e. all occupied states are
involved in transitions.
Therefore, if $V<v_F/2$ we get for the drag current
\begin{equation}\label{qc05}
J_d\propto\,\int_{-p_F}^{-p_F+2mV}dp\frac{K_0^2[2(mV-p)d/\hbar]}{mV-p}=\int_{p_F-mV}^{p_F+mV}dp\frac{K_0^2(2pd/\hbar)}{p}
\end{equation}
and we see that increase in $V$ decreases the minimal transferred
momentum and increases the effective Coulomb interaction
$K_0(2pd/\hbar)$. If $V>v_F/2$ we have
\begin{equation}\label{qc06}
J_d\propto\,\int_{-p_F}^{0}dp\frac{K_0^2\left[2(mV-p)d/\hbar\right]}{mV-p}=\int_{mV}^{p_F+mV}dp\frac{K_0^2(2pd/\hbar)}{p}
\end{equation}
and increase of $V$ results in a decrease of the drag current. These
equations provide an adequate description of the drag current dependence on the ion beam
velocity as can be readily seen in our quantitative approach below.

\section{Interaction of ion beam with electrons of nanostructure}

For simplicity, we assume the width of the beam to be constant (actually
it may slightly vary in the course of beam propagation). Then one can write the
distribution of the ions within the beam as
\begin{equation}
F_{\bf P}=N(2\pi\hbar)^3\delta(P_x)\delta(P_y)\delta(P_z-P),
\label{dist_funct}
\end{equation}
$N$ being the ion concentration.

The collision term of the Boltzmann equation for 1D electrons and
3D ions is in the Born approximation
\begin{eqnarray}
\left[\frac{{\partial f}_{pn}}{\partial t} \right]_{\rm
coll}\equiv I\{f, F\}=\int \frac{{\cal V}_{\bf
R}d^3P}{(2\pi\hbar)^3}\int\frac{{\cal V}_{\bf
R}d^3q}{(2\pi\hbar)^3}\frac{2\pi}\hbar\left|\langle{
p,n},{\bf P} |U|{{p+q_z,n},{\bf P-q}}\rangle\right|^2\nonumber\\
\times\delta(\epsilon_{n{p}}+E_{\bf P}-\epsilon_{n,{p+q_z}}-E_{\bf
P-q})\left[f_{n{p}}(1- f_{n,{p+q_z}})F_{\bf
P}-f_{np+q_z}(1-f_{np})F_{{\bf P}-{\bf q}}\right]
 \label{sc_int}
\end{eqnarray}
where
\begin{equation}
\epsilon_n(p)=\epsilon_n(0)+p^2/2m
 \label{7e}
 \end{equation}
Here $n$ is the number of the channel, i.e. of the mini-band of
1D transverse quantization (according to the assumption
made above this number does not change in the course of electron
transitions), $\bf q$ is the transferred (quasi)momentum, $M$ is
the mass of an ion, $m$ is the effective mass of conduction
electron and
\begin{equation}
U=\frac{2}{1+\kappa}\frac{ee_{I}}{|{\bf R-r}|} \label{002}
\end{equation}
describes the Coulomb interaction of the ion with charge $e_I$ and
electron in the wire, $\kappa$ being the dielectric susceptibility of
the wire. For the matrix element in Eq.(\ref{sc_int}) we have
\begin{equation}
\langle{p,n,{\bf P}} |U|p+q_z,n,{\bf P}-{\bf q}\rangle= \int_{\cal
V_{\bf r}} d^3r\int_{\cal V_{\bf R}} d^3R\psi^{\ast}_{n}({\bf
r}_{\perp})\Psi^*_{\bf P}\frac{2ee_{I}}{L(1+\kappa)|{\bf r-R}|}
\psi_{n}({\bf r}_{\perp})\Psi_{{\bf P-q}}e^{iq_zz/\hbar}.
 \label{7}
 \end{equation}
Since
\begin{equation}
\int  \frac{dZ dz}{L|{\bf r-R}|} e^{iq_z(z-Z)/\hbar}=2K_0(|q_z||{
\Delta{\bf r_{\perp}}}|/\hbar)
 \label{8}
 \end{equation}
where
$$
|\Delta{\bf r_{\perp}}|\equiv\sqrt{(x-X)^2+(y-Y)^2}
$$
we can write
\begin{eqnarray}
\langle{p,n,{\bf P}}|U|p+q_z,n,{\bf P}-{\bf
q}\rangle=\nonumber\\\frac{4 e e_I}{(1+\kappa){\cal V}_{\bf R}}\int
d{\bf R}_{\perp}\int d{\bf r}_{\perp}|\psi_n({\bf
r}_{\perp})|^2{e^{-i{\bf q}_{\perp}{\bf
R}_{\perp}/\hbar}}K_0(|q_z||\Delta{\bf r}_{\perp}|/\hbar).
 \label{104}
\end{eqnarray}

The Boltzmann equation for electrons is
\begin{equation}
v\frac{\partial f_{np}}{\partial z}=-\left[\frac{{\partial f}_{n
p}}{\partial t} \right]_{\rm coll},
 \label{003}
\end{equation}
where
$$
v=\frac{d\epsilon_{pn}}{dp}=\frac pm
$$
is the electron velocity.

To calculate the current in the wire, we iterate the Boltzmann
equation for the electrons of the wire in the term describing
collisions between electrons of the wire and ions. In the zeroth
approximation one can choose the electron distribution function in
the collision term to be equilibrium one. In what follows
$f_{np}\equiv\,f_F(\epsilon_{np}-\mu)$ will be implied, where
$f_F$ is the Fermi distribution function while
$\mu$ is the Fermi level. The first iteration of Eq.(\ref{003})
gives for the nonequilibrium part of the distribution function
\begin{equation}
\Delta f_{n{p}}=-\left(z\pm\frac L2 \right)\frac1{v_{np}}I\{f, F\}
 \label{004}
 \end{equation}
 for $p>0$ and
$p<0$, respectively. Here $I\{f, F\}$ is a shorthand notation for the
collision term. Using the particle conservation property of
the scattering integral
\begin{equation}
\sum_n\int dp I\{f, F\}=0
 \label{004a}
 \end{equation}
we get for the drag current $J_d$ (cf. with Ref. \onlinecite{GM1})
\begin{equation}
J_d=-2eL\sum_n\int_0^{\infty}\frac{dp}{2\pi\hbar} I\{f, F\}.
 \label{005}
 \end{equation}
With the distribution function given by Eq.(\ref{dist_funct}) we
have
\begin{eqnarray}
J_d=-2eN{\cal V}_{\bf
R}^2L\sum_n\int_0^{\infty}\frac{dp}{2\pi\hbar}
\int\frac{d^3q}{(2\pi\hbar)^3}\frac{2\pi}{\hbar}\left\{\left|\langle{
p,n},P{\bf e}_z |U|{{p+q_z,n},P{\bf e}_z-{\bf
q}}\rangle\right|^2\right.\\\nonumber
\times\delta(\epsilon_{n{p}}+E_{P{\bf
e}_z}-\epsilon_{n,{p+q_z}}-E_{P{\bf e}_z-{\bf q}})f_{n{p}}(1-
f_{n,{p+q_z}})-\\\nonumber - \left|\langle{p,n},P{\bf e}_z+{\bf
q} |U|{{p+q_z,n},P{\bf e}_z}\rangle\right|^2
\left.\delta(\epsilon_{n{p}}+E_{P{\bf e}_z+{\bf
q}}-\epsilon_{n,{p+q_z}}-E_{P{\bf e}_z})f_{n{p+q_z}}(1-
f_{n,{p}})\right\}
 \label{100}
\end{eqnarray}
where ${\bf e}_z$ is the unit vector along the $z$ axis.
In the first term under the integral we change ${\bf
q}\rightarrow\,-{\bf q}$ and shift the integration variable $p$ by
$q_z$, then
\begin{eqnarray}
J_d=-2eN{\cal V}_{\bf R}^2L\sum_n
\int\frac{d^3q}{(2\pi\hbar)^3}\int_{0}^{q_z}\frac{dp}{2\pi\hbar}\frac{2\pi}{\hbar}
\left|\langle{p,n,P{\bf e}_z+{\bf q}}|U|p+q_z,n,P{\bf
e}_z\rangle\right|^2\\\nonumber
\times\delta(\epsilon_{n{p}}+E_{P{\bf e}_z+{\bf
q}}-\epsilon_{n,{p-q_z}}-E_{P{\bf e}_z})f_{n{p-q_z}}(1- f_{n,{p}}),
 \label{102}
\end{eqnarray}
so that the drag current is
\begin{eqnarray*}
J_d=-J_0\frac{2MS_{\bf R}}{m\pi^2\hbar^2}\sum_n \int
dq_z\int_0^{q_z} dp f_{n{p-q_z}}(1- f_{n,{p}})\int d{\bf
q}_{\perp}
g({\bf q}_{\perp},|q_z|)\nonumber\\
\times\delta\left[q_{\perp}^2-q_z^2(M/m-1)+2q_z(Mp/m+P)\right].
\end{eqnarray*}
Here we introduced
\begin{eqnarray*}
J_0=\frac{e(2ee_I)^2LNmS_{\bf R}}{(1+\kappa)^2\pi \hbar^3}
\end{eqnarray*}
and a dimensionless quantity $g({\bf q}_{\perp},|q_z|)$ according
to
\begin{equation}
S_{\bf R}^2g({\bf q}_{\perp},|q_z|)=\left|\int d{\bf
R}_{\perp}d{\bf r}_{\perp}{e^{-i{\bf q}_{\perp}{\bf
R}_{\perp}/\hbar}}|\psi_n({\bf r}_{\perp})|^2K_0(|q_z||\Delta {\bf
r}_{\perp}|/\hbar)\right|^2
\end{equation}
where $S_{\bf R}$ is the cross section area of the ion beam.

We get
\begin{eqnarray}
J_d=J_0\frac{2MS_{\bf R}}{m\pi^2\hbar^2}\sum_n \int_0^{\infty}
dq_z\int_0^{q_z} dp f_{n{p-q_z}}(1- f_{n,{p}})\int d{\bf
q}_{\perp}
g({\bf q}_{\perp},q_z)\nonumber\\
\times\left\{\delta\left[q_{\perp}^2-q_z^2(M/m-1)+2q_z(Mp/m-P)\right]\right.\nonumber\\
\left.-\delta\left[q_{\perp}^2-q_z^2(M/m-1)+2q_z(Mp/m+P)\right]\right\}.
\label{dr_curr}
\end{eqnarray}
\subsection{Linear response}
In the linear response regime
\begin{equation}
 V\,\ll\,T/p_{nF}\quad\mbox{where}\quad p_{nF}=\sqrt{2m[\mu-\epsilon_n(0)]}
\label{59a}
\end{equation}
the difference of $\delta$-functions in Eq.(\ref{dr_curr}) can be
expanded as (we again take into account that $M/m\,\gg\,1$)
\begin{eqnarray*}
\delta\left[q_{\perp}^2-q_z^2M/m+2q_z(Mp/m+P)\right]-\delta\left[q_{\perp}^2-q_z^2M/m+2q_z(Mp/m-P)\right]\\=
\frac{m}{2M|q_z|}\left\{\delta\left[mq_{\perp}^2/2Mq_z-q_z/2+p+mV\right]-\delta\left[mq_{\perp}^2/2Mq_z-q_z/2+p-mV\right]\right\}\\=
mV\frac{m}{M|q_z|}\frac{\partial}{\partial
p}\delta\left[mq_{\perp}^2/2Mq_z-q_z/2+p\right].
\end{eqnarray*}
Then integration by parts gives
\begin{eqnarray}
J_d=J_0\frac{2S_{\bf R}}{\pi^2 \hbar^2}mV\sum_n \int_0^{\infty}
\frac{dq_z}{q_z}\int_0^{q_z} dp \int d{\bf q}_{\perp} g({\bf
q}_{\perp},q_z)\nonumber\\\times\delta\left[mq_{\perp}^2/2Mq_z-q_z/2+p\right]\frac{\partial}{\partial
p}f_{n{p-q_z}}(1- f_{n,{p}}). \label{dr_curr01}
\end{eqnarray}
Using
\begin{equation}\label{diff_df}
\frac{\partial}{\partial p}f_{n{p-q_z}}(1- f_{n,{p}})=(1-
f_{n,{p}})\delta(q_z-p-p_{nF})+f_{n{p-q_z}}\delta(p-p_{nF})
\end{equation}
we have
\begin{eqnarray}
J_d=J_0\frac{4S_{\bf R}}{\pi^2 \hbar^2}mV\sum_n
\int_{p_{nF}}^{\infty} dq_z \int d{\bf q}_{\perp} g({\bf
q}_{\perp},q_z)\nonumber\\\times\left\{\delta\left[q^2_z-2p_{nF}q_z+mq_{\perp}^2/M\right](1-
f_{n,q_z-p_{nF}})+\delta\left[q^2_z-2p_{nF}q_z-mq_{\perp}^2/M\right]
f_{n,p_{nF}-q_z} \right\}.\label{dr_curr01a}
\end{eqnarray}
Eliminating the $\delta$-functions we get
\begin{eqnarray*}
J_d=J_0\frac{2S_{\bf R}}{\pi^2 \hbar^2}mV\sum_n \int d{\bf
q}_{\perp} \left\{g({\bf q}_{\perp},p_{nF}+p_1)\frac{(1-
f_{n,p_1})}{p_1}+g({\bf q}_{\perp},p_{nF}+p_2)\frac{
f_{n,p_2}}{p_2}\right\},
\end{eqnarray*}
where $p_1=\sqrt{p_{nF}^2-mq_{\perp}^2/M}$ and
$p_2=\sqrt{p_{nF}^2+mq_{\perp}^2/M}$. The expression for $J_d$ can be
simplified as follows
\begin{equation}
J_d=J_0\frac{4S_{\bf R}}{\pi^2 \hbar^2}mV\sum_n
\frac{1}{p_{nF}}\int d{\bf q}_{\perp}\frac{g({\bf
q}_{\perp},2p_{nF})}{e^{q_{\perp}^2/2MT}+1}. \label{5}
\end{equation}

If the ion beam cross section is of a circular form with radius
$a$ we have
\begin{equation}
g({\bf q}_{\perp},2p_{nF})=\left(\frac{2\hbar
J_1(aq_{\perp}/\hbar)}{aq_{\perp}}\right)^2K_0^2(2p_{nF}d/\hbar),
\label{7a1}
\end{equation}
where $J_1(x)$ is the Bessel function of the first order and $d$
is the distance between the ion flux and the wire.

Below we will discuss in more detail a special case where
$g({\bf q}_{\perp},2p_{nF})$ does not depend on
${\bf q}_{\perp}$. For instance, this is the case provided
\begin{equation}
\sqrt{MT}\,\ll\,\hbar/a.
\label{7a14}
\end{equation}
Then for ${\cal N}=1$ we get
\begin{equation}
J_d=J_0\frac{4\ln{(4)}a^2}{\hbar^2}(MT)\frac{V}{v_F}
K_0^2(2p_{F}d/\hbar),\label{8a}
\end{equation}
where $v_F=p_{F}/m$ is the Fermi velocity and
\begin{equation}
J_0=\frac{e(2ee_I)^2LNm a^2}{(1+\kappa)^2 \hbar^3}.
\end{equation}

For the opposite case where
\begin{equation}
\sqrt{MT}\,\gg\,\hbar/a \label{7f}
\end{equation}
the drag does not depend on temperature. For the values
$M=10^{-22}$g (Ga), $T=4$K, $a=10^{-5}$cm this inequality can be
easily satisfied. Then we have
\begin{equation}
J_d=J_0\frac{8V}{v_F}
K_0^2\left(\frac{2p_{F}d}{\hbar}\right).\label{8a1}
\end{equation}

It is interesting to calculate the ratio $J_d/J_I$ for this case
\begin{equation}
\frac{J_d}{J_I}=\frac e{e_I}\frac{32(ee_I)^2Lm}{(1+\kappa)^2
\pi\hbar^3v_F} K_0^2\left(\frac{2p_{F}d}{\hbar}\right).\label{8a2}
\end{equation}
Here one can use for $K_0(s)$ equations (\ref{11e}) and (\ref{11f}).

For an estimate we assume the following values $L=10^{-4}$cm,
$m=7\cdot10^{-29}$g, $v_F=2\cdot10^7$cm/s, $\kappa=10$,
$p_Fd/\hbar=2$, so that $K_0^2(2p_Fd/\hbar)=1.3\cdot10^{-4}$. Then
for $J_I=10^{-8}$A one gets $J_d=2\cdot10^{-9}$A and the
corresponding drag voltage ${\cal V}_d$ is about
\begin{equation}
{\cal V}_d=20\,\,\mu\mbox{V}. \label{11j}
\end{equation}

Naturally, if $J_I$ goes up ${\cal V}_d$ also goes up in
proportion to $J_I$.

\subsection{Nonlinear case}
We consider the simplest case of low temperatures assuming that
\begin{equation}
V\,\gg\,T/p_F.
\label{59b}
\end{equation}
In our further calculation we will assume $T=0$,
then the integration due to the Fermi functions in
Eq.(\ref{dr_curr}) is restricted and we get (the first or the
second $\delta$-function contributes
 for $V>0$ and $V<0$ respectively, so that the drag current
changes its sign with $V$ as it should)
\begin{eqnarray*}
J_d=J_0\frac{a^2}{\pi\hbar^2}\sum_n \left(\int_{p_{nF}}^{2p_{nF}}
\frac{dq_z}{q_z}\int_{p_{nF}}^{q_z} dp +\int_{2p_{nF}}^{\infty}
\frac{dq_z}{q_z}\int_{q_z-p_{nF}}^{q_z} dp\right)\\
\times\int d{\bf q}_{\perp} g({\bf q}_{\perp},q_z)
\delta\left[p-mV-q_z(1-m/M)/2+mq_{\perp}^2/2q_zM\right].
\end{eqnarray*}

The result valid for
\begin{equation}
V<v_F/2
\label{59bx}
\end{equation}
is
\begin{equation}
J_d=J_0\frac{a^2}{\pi\hbar^2}\sum_n\int d{\bf
q}_{\perp}\Theta[4Pp_{nF}-q_{\perp}^2]\int_{p_{-}}^{p_{+}}\,\frac{dq_z}{q_z}g({\bf
q}_{\perp},q_z)\label{59j}
\end{equation}
where $\Theta$ is the step function and
$$p_{\pm}=p_{nF}\pm mV+\sqrt{(p_{nF}\pm mV)^2\mp
mq_{\perp}^2/M}$$
(other cases are considered in Appendix A).

For $V\ll\,v_{nF}$ the integration variable $q_z$ is in the
vicinity of $2p_{nF}$ and we have
\begin{eqnarray}
J_d=J_0\frac{a^2}{2\pi\hbar^2}\sum_n\int d{\bf
q}_{\perp}\Theta[4Pp_{nF}-q_{\perp}^2]\left(4\frac{mV}{p_{nF}}-\frac{m}{M}\frac{q_{\perp}^2}{p_{nF}^2}\right)
g({\bf q}_{\perp},2p_{nF}). \label{59ab}
\end{eqnarray}
Eq.(\ref{59ab}) substantially simplifies provided $g({\bf q}_{\perp},2p_{nF})$ does not depend on ${\bf
q}_{\perp}$; this is the case provided the ion flux cross section
characteristic width $a$ obeys the inequality
\begin{equation}
\sqrt{Pp_{nF}}a/\hbar\,\ll\,1.
\label{9t}
\end{equation}
Then
\begin{equation}
J_d=J_1\sum_n g(2p_{nF}),
\label{9s}
\end{equation}
where
\begin{equation}
J_1=J_0\frac{(2mV)^2a^2}{2\hbar^2}\frac{M}{m}.
\end{equation}

This expression is valid for $V\,>\,0$, i.e. when
the ion flux is directed "to the right". Then the momentum
transferred to the electron system in the wire is also
directed to the right and the current (since $e\,<\,0$) flows in
the opposite direction regardless of the sign of dragging ion
charge.

Assuming that the distance $d$ between the ion flux and the wire
is much bigger than the characteristic cross section length of the
wire and the flux we can write
\begin{equation}
g({\bf q}_{\perp},2p_{nF})\simeq\,K_0^2(2p_{nF}d/\hbar), \label{9}
\end{equation}
and
\begin{equation}
J_d=J_1\sum_nK_0^2(2p_{nF}d/\hbar). \label{10}
\end{equation}
Using for the function $K_0$ the approximate equation (\ref{11f})
we get
\begin{equation}
J_d=J_1\sum_n \frac{1}{k_{nF}d}e^{-4k_{nF}d},
\label{11}
\end{equation}
where $k_{nF}=p_{nF}/\hbar$.

In the case $a\sqrt{Pp_{nF}}/\hbar\,\gg\,1$ the drag current is
linear in $V$
\begin{equation}
J_d=4J_0MV\sum_n\frac{1}{p_{nF}}K_0^2(2p_{nF}d/\hbar).
\label{9r_mur}
\end{equation}

\begin{figure}[htb]\label{zavisimotskorosti}
\begin{center}
\includegraphics[width=10 cm]{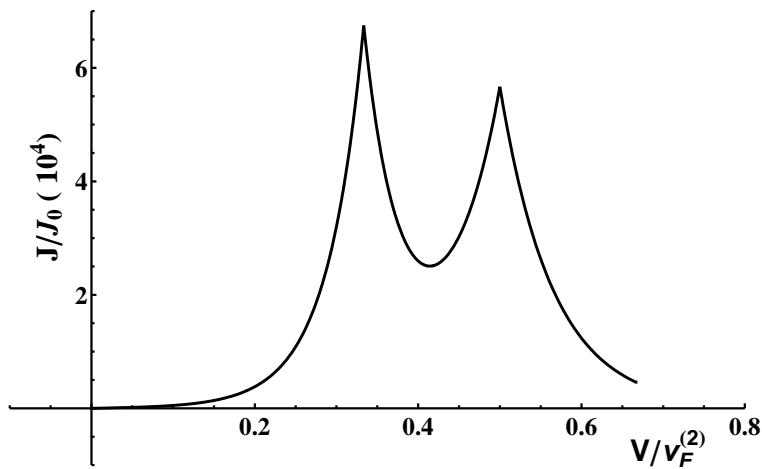}\\
\end{center}
\caption{Drag current dependence on the velocity of the ion beam.
We take $v_{F}^{(2)}=kv_{F}^{(1)}$ and $k=3/2$. The first peak
corresponds to $V/v_{F}^{(2)}=1/2k$ (i.e. $V/v_{F}^{(1)}=1/2$) and
the second peak corresponds to $V/v_{F}^{(2)}=1/2$. Here
$J_0=e(2ee_I)^2LNma^2/(1+\kappa)^2\hbar^3$.}
\end{figure}

\section{Conclusive remarks}
We have developed a theory of Coulomb drag of electrons in 1D
ballistic nanostructure by an ion beam. This provides an example
of drag of quasiparticles of the nanostructure by particles of
the beam. It is worthwhile to mention that such a beam may consist
not only of heavy ions but also of electrons. The free electron mass
is usually bigger than the effective mass of conduction electrons, so
that the adopted approximations of our calculation, $M\gg m$, may remain
valid in this case too.

The experimental setup should permit one to vary the velocity $V$
within rather wide limits. We see however that to achieve a large
drag effect one should choose the value of $V$ near to $v_{Fn}/2$
(see Fig.~2). This means in particular that
the ion beam drag may be a useful instrument for nanostructure
spectroscopy: it may make it possible to measure with appropriate
accuracy the Fermi velocities $v_{Fn}$ in each channel $n$.

\appendix \label{A1}
\section{Evaluation of the drag current for various ratios of
$\alpha=V/v_F$}

We introduce dimensionless parameters $$\alpha=mV/p_{nF}=V/v_{nF}$$
and $$b=mq_{\perp}^2/p_{nF}^2M$$ and write $q$ instead of
$q(1-m/M)$ in the argument of the $\delta$-function
\begin{eqnarray}
J_d=J_0\sum_n
\left(\frac{p_{nF}a}{\hbar}\right)^2\frac{1}{\pi}\left(\int_{1}^{2}
dq\int_{1}^q dp +\int_{2}^{\infty}
dq\int_{q-1}^q dp\right)\\\nonumber
\times\frac{1}{q}\int d{\bf q}_{\perp} g(p_{nF}{\bf
q}_{\perp},p_{nF}q)\delta\left[p-\alpha-q/2+b/2q\right]\label{A59}
\end{eqnarray}
where $$J_0=e(2ee_I)^2LNma^2/(1+\kappa)^2\hbar^3.$$ For $\alpha<1/2$
we get
\begin{eqnarray}
J_d=J_0\sum_n \left(\frac{p_{nF}a}{\hbar}\right)^2\frac{1}{\pi}\int
d{\bf q}_{\perp}\Theta[4\alpha-b]
\int_{A_+(-\alpha,-b)}^{A_+(\alpha,b)}dq\frac{g(p_{nF}{\bf
q}_{\perp},p_{nF}q)}{q} \label{A60}
\end{eqnarray}
where $A_{\pm}(\alpha,b)=1+\alpha\pm\,\sqrt{(1+\alpha)^2-b}$.

If $1/2<\alpha<1$ we get
\begin{eqnarray}
J_d=J_0\sum_n \left(\frac{p_{nF}a}{\hbar}\right)^2\frac{1}{\pi}\int
d{\bf
q}_{\perp}\left\{\Theta[2\alpha-1-b]\int_{A_+(\alpha-1,b)}^{A_+(\alpha,b)}dq\right.\\\nonumber
\left.+\Theta[b-2\alpha+1]\Theta[4\alpha-b]
\int_{A_+(-\alpha,-b)}^{A_+(\alpha,b)}dq\right\}\frac{g(p_{nF}{\bf
q}_{\perp},p_{nF}q)}{q}.\label{A61}
\end{eqnarray}
We will not give here explicit expressions for larger values of
$V/v_F$ but rather present the simple expression for the drag
current valid for $\hbar/a\,\ll\,mV\sqrt{M/m}$
\begin{eqnarray}
J_d=4J_0\int_{\alpha}^{\infty}\,dz\frac{K_0^2(2p_Fzd/\hbar)}{z}\left(\frac{1}{e^{([z-\alpha]^2-1)p_F^2/2mT}+1}
-\frac{1}{e^{([z+\alpha]^2-1)p_F^2/2mT}+1}\right).\label{A62}
\end{eqnarray}
This expression is reduced to Eq.(\ref{8a1}) and Eq.(\ref{9r_mur})
in the corresponding limiting cases. For $mV\,\gg\,T/v_F$ the
difference of the Fermi functions restricts the integration region
so that for $\alpha<1/2$ we have
\begin{eqnarray}
J_d=4J_0\int_{1-\alpha}^{1+\alpha}\,dz\frac{K_0^2(2p_Fzd/\hbar)}{z}\label{A63}
\end{eqnarray}
and
\begin{eqnarray}
J_d=4J_0\int_{\alpha}^{1+\alpha}\,dz\frac{K_0^2(2p_Fzd/\hbar)}{z}\label{A64}
\end{eqnarray}
for $\alpha>1/2$. The drag current calculated according to these
simple  formulas practically coincides with that calculated from
the exact expressions and presented in
Fig.~2 for the case ${\cal N}=2$.

\section{Preferred velocity of the beam}
Let us differentiate  Eq.(\ref{dr_curr}) with respect to $mV$ and
determine the sign of derivative. We get under the sign of integral the
following sum of $\delta$-functions
\begin{eqnarray}
-\frac{d}{dp}\left\{\delta\left[q_{\perp}^2m/M-q_z^2+2q_z(p-mV)\right]
+\delta\left[q_{\perp}^2m/M-q_z^2+2q_z(p+mV)\right]\right\}.
\end{eqnarray}
We integrate over $p$ by parts and get (we denote the ratio
$J_d\pi^2\hbar^2/(2J_0S_{\bf R}m)$ by $j$)
\begin{eqnarray}\label{vtorur}
\frac{dj}{dV}=\int_0^{\infty} dq_z\int\,d{\bf q}_{\perp} g({\bf
q}_{\perp},q_z)\int_0^{q_z} dp
\left\{\delta\left[q_{\perp}^2m/M-q_z^2+2q_z(p-mV)\right]\right.\\
+\left.\delta\left[q_{\perp}^2m/M-q_z^2+2q_z(p+mV)\right]\right\}
\frac{d}{dp}\left[f_{n{p-q_z}}(1- f_{n,{p}})\right]\nonumber\\
-\int_{0}^{\infty}dq_z\int\,d{\bf q}_{\perp} g({\bf
q}_{\perp},q_z)(1-f_{q_z})\left\{\delta\left[q_{\perp}^2m/M+q_z^2-2q_zmV\right]\right.\nonumber\\
+\left.\delta\left[q_{\perp}^2m/M+q_z^2+2q_zmV\right]\right\}\nonumber
\end{eqnarray}
We take into account the strong Fermi degeneracy of the electron system
so that $1-f_0=0,\,f_0=1$. Using Eq.(\ref{diff_df}) we get
\begin{eqnarray}
\frac{dj}{dV}=\int_{2p_{nF}}^{\infty} dq_z\int\,d{\bf q}_{\perp}
g({\bf
q}_{\perp},q_z)\left\{\delta\left[q_{\perp}^2m/M+q_z^2-2q_z(p_{nF}+mV)\right]+\right.\nonumber\\
\left.\delta\left[q_{\perp}^2m/M+q_z^2-2q_z(p_{nF}-mV)\right]\right\}\nonumber\\
-\int_{p_{nF}}^{\infty}dq_z\int\,d{\bf q}_{\perp} g({\bf
q}_{\perp},q_z)\delta\left[q_{\perp}^2m/M+q_z^2-2q_zmV\right]\label{A66}
\end{eqnarray}
We again use $f_{q_z-p_{nF}}=1$ for $q_z>p_{nF}$ and take into
account that $V>0$ so that the last $\delta$-function in
Eq.(\ref{vtorur}) does not contribute. The second
$\delta$-function in the first integral in the previous expression
does not contribute as well and we arrive at
$$
\frac{dj}{dV}=I_{+}-I_{-},
$$
where we introduce notations for the positive and negative
integrals
\begin{eqnarray}
I_{+}=\int d{\bf q}_{\perp} \int_{2p_{nF}}^{\infty} dq_zg({\bf
q}_{\perp},q_z)\delta\left[q_{\perp}^2m/M+q_z^2-2q_z(p_{nF}+mV)\right]
\end{eqnarray}
\begin{eqnarray}
I_{-}=\int d{\bf q}_{\perp} \int_{p_{nF}}^{\infty}dq_zg({\bf
q}_{\perp},q_z)\delta\left[q_{\perp}^2m/M+q_z^2-2q_zmV\right]
\end{eqnarray}
If $V/v_{nF}<1/2$ we get
\begin{eqnarray}
I_{+}=\int_{q_{\perp}^2<4Mp_{nF}V} d{\bf q}_{\perp} \frac{g({\bf
q}_{\perp},q_1)}{2\sqrt{(p_{nF}+mV)^2-mq_{\perp}^2/M}}
\end{eqnarray}
where $q_1=p_{nF}+mV+\sqrt{(p_{nF}+mV)^2-mq_{\perp}^2/M}$ and
$I_{-}=0$ and the drag current is an increasing function of the
beam velocity $V$.

The integral $I_{-}$ has nonzero values only if $V/v_{nF}>1/2$. If
$V/v_{nF}<1$ we have for this integral
\begin{eqnarray}
I_{-}=\int_{q_{\perp}^2<Mp_{nF}(2mV-p_{nF})/m} d{\bf q}_{\perp}
\frac{g({\bf q}_{\perp},q_2)}{2\sqrt{(mV)^2-mq_{\perp}^2/M}}
\end{eqnarray}
where $q_2=mV+\sqrt{(mV)^2-mq_{\perp}^2/M}$. $I_{-}$ in this
region becomes larger than $I_{+}$ (the latter being practically
zero due to exponential dependence on $q_1$) and the drag current
turns into decreasing function of the velocity $V$.

If $V/v_{nF}>1$
\begin{eqnarray}
I_{-}=\int_{q_{\perp}^2<Mp_{nF}V} d{\bf q}_{\perp} \frac{g({\bf
q}_{\perp},q_2)}{2\sqrt{(mV)^2-mq_{\perp}^2/M}}\\
-\int_{q_{\perp}^2>Mp_{nF}(mV-p_{nF})/m} d{\bf q}_{\perp}
\frac{g({\bf q}_{\perp},q_3)}{2\sqrt{(mV)^2-mq_{\perp}^2/M}}
\end{eqnarray}
where $q_3=mV-\sqrt{(mV)^2-mq_{\perp}^2/M}$.

Therefore we see that the drag current has a maximum as a function
of the beam velocity in the vicinity of $V=v_{nF}/2$.

\end{document}